\documentclass[conference]{IEEEtran}
\IEEEoverridecommandlockouts
\usepackage{cite}
\usepackage{url}
\usepackage{amsmath,amssymb,amsfonts}
\usepackage{algorithmic}
\usepackage{graphicx}
\usepackage{textcomp}
\usepackage[table]{xcolor}
\usepackage{colortbl}
\usepackage{authblk}
\usepackage[bookmarks=false]{hyperref}

\def\BibTeX{{\rm B\kern-.05em{\sc i\kern-.025em b}\kern-.08em
    T\kern-.1667em\lower.7ex\hbox{E}\kern-.125emX}}
\begin{document}

\title{The Performance of Machine and Deep Learning Classifiers in Detecting Zero-Day Vulnerabilities}

\author[1]{Faranak Abri}
\author[2]{Sima Siami-Namini}
\author[3]{Mahdi Adl Khanghah}
\author[3]{Fahimeh Mirza Soltani}
\author[1]{Akbar Siami Namin}

\affil[1]{\footnotesize Department of Computer Science, Texas Tech University, USA}
\affil[2]{\footnotesize Department of Mathematics and Statistics, Texas Tech University University, USA}
\affil[3]{\footnotesize Department of Computer Science, University of Debrecen, Hungary}
\affil[ ]{\footnotesize \{faranak.abri, sima.siami-namini, akbar.namin\}@ttu.edu; \{adl.mahdi1365, soltani.fahimeh1364\}@gmail.com}

	
%

\maketitle

\begin{abstract}
The detection of zero-day attacks and vulnerabilities is a challenging problem. It is of utmost importance for  network administrators to identify them with high accuracy. The higher the accuracy is, the more robust the defense mechanism will be. In an ideal scenario (i.e., 100\% accuracy) the system can detect zero-day malware without being concerned about mistakenly tagging benign files as malware or enabling disruptive malicious code running as none-malicious ones. This paper investigates different machine learning algorithms to find out how well they can detect zero-day malware. Through the examination of 34 machine/deep learning classifiers, we found that the random forest classifier offered the best accuracy. The paper poses several research questions regarding the performance of machine and deep learning algorithms when detecting zero-day malware with zero rates for false positive and false negative.

\end{abstract}


\section{Introduction}
\label{sec:introduction}

Malware is a malicious application, specifically developed to perform malicious activities including disruption, damaging the underlying computer system and data, or gaining authorized access to the computer systems. A common approach to identify malicious applications is through signature-based matching, where the profile of a suspicious application is compared against the reported ones. The popular intrusion detection systems often employ such pattern-matching techniques and tools such as YARA \cite{YARA} to implement rule-based detection mechanisms. These pattern matching approaches to the malware detection problem, however, are static and thus not capable of detecting all types of malware. Broadly speaking, pattern and signature-based malware detection mechanisms have the ability of detecting only a small subset of all classes of malware software and thus these approaches are less effective in detecting more sophisticated, obfuscated, unknown, or newly developed malware software. A special case of interest is zero-day, a type of malware with no history or clear remediation strategy. 

The software vulnerability, which is unknown to system administrators and thus there is no known security patches or remedies to confront it, is called a zero-day vulnerability. Due to the absent of a structured remediation strategy and mitigation plan, such vulnerabilities are exploitable by adversaries.

The signature-based approaches to malware detection cannot be utilized to catch zero-day malware applications. As a result, instead of syntactical approaches (i.e., signature-based) analytical techniques should be employed. Approaches such as behavioral-based \cite{Xie:2010}, data mining-based \cite{Ye:2017}, learning-based \cite{Rieck:2008}, machine learning \cite{ML}, and more notably deep learning-based \cite{DMIN}  have already been proposed to detect malware.

The conventional machine learning algorithms developed for addressing clustering problems employ various forms of classification techniques to group a given dataset into distinct sub-classes. For instance, regression-based approaches partition the given data into two or more clusters and then fit multiple linear regression lines on the data within each cluster; whereas, the Support Vector Machine (SVM) algorithms intend to partition the given data into clusters by fitting a hyper-plane into the data with acceptable margins. Furthermore, ensemble learning-based approaches such as random forest build multiple decision trees and then take the majority votes  of these decision trees and then perform clustering.  

As an extension to conventional machine learning algorithms, the deep learning-based approaches  \cite{Siami-NaminiTN18, Tavakoli19, ChatterjeeN19} take the clustering problem into a deeper level by exploring the hidden relationships between the features of the dataset and then building a hierarchical model to perform clustering. While the time complexity of building such models is high, the deep exploration of data helps in building more accurate clustering. 

Given the recent advancement in deep learning-based approaches to model prediction problems, a major challenging question is whether these learning-based techniques and in particular machine and deep learning algorithms are effective in detecting zero-day malicious software. This paper investigates the performance of the conventional machine learning techniques in comparison with their counterparts, deep learning-based algorithms. The goal is to empirically observe whether these learning-based approaches are a suitable modeling approach for detecting zero-day vulnerabilities. The paper conducts several experiments with respect to four learning-based clustering techniques:

\begin{enumerate}
    \item Conventional machine learning
    \item Simple neural network learner (i.e., Single layer)
    \item Deep learning with multiple layers
    \item Deep(er) learning with multiple hidden layers  
\end{enumerate}

While it is expected that these learning-based algorithms demonstrate acceptable performance and accuracy, it is crucial that the malware detection systems learned and developed for catching zero-day vulnerabilities do not label legitimate applications as malware (i.e., zero false positive), and do not label malicious applications as legitimate (i.e., zero false negative). An elevated false positive and false negative rates will cause either eliminating important files or enabling malicious applications to be installed and then activated on the underlying operating systems and thus it leads to serious loss or damage to the assets. Hence, it is of utmost importance to achieve extremely high accuracy and extremely low false positive/negative predictions when building the model. 

This paper investigates the performance of the aforementioned learning-based classes of algorithms for a given dataset and measures the accuracy of the models built using popular cross validation approaches ($k$-fold). The key contributions of this paper are as follows:
\begin{enumerate}
    \item Conduct several experimental studies to compare the performance of various types of machine and deep learning algorithms in detecting zero-day vulnerabilities.
    \item Report that no individual learning algorithm was able to achieve 100\% accuracy when fitting the models.
    \item Report that the conventional machine learning algorithms are as good as deep learning-based models.
    \item Report that random forest outperforms other machine and deep learning algorithms in providing the best models for detecting zero-day vulnerabilities. 
\end{enumerate}

The paper is organized as follows: Section \ref{sec:researchquestion} highlights the research questions to address. A background on zero-day malware and vulnerability detection is presented in Section \ref{sec:background}. The experimental setup is provided in Section \ref{sec:setup}. Section \ref{sec:classifiers} lists the classifiers along with their parameters studied in this paper. Section \ref{sec:results} reports the results and compares the performance of different machine and deep learning algorithms. Section \ref{sec:discussion} discusses some implications observed while analyzing the experimental data. Section \ref{sec:conclusion} concludes the paper and highlights the future research directions.

\section{Challenges and Research Questions}
\label{sec:researchquestion}

This paper empirically investigates whether it is possible to train learning-based algorithms and then employ the developed model to predict whether a zero-day vulnerability or malware with no prior historical data can be detected. More specifically, the experiments conducted in this paper intend to address the following research questions:

\begin{description}
\item[\bf RQ. 1] Does standardization of data help in enhancing the accuracy? 
\item[\bf RQ. 2] Is it possible to train a machine/deep learning algorithm and achieve 100\% accuracy for prediction and thus reduce the false positive/negative ratios to zero for detecting zero-day vulnerabilities/malware? 
\item[\bf RQ. 3] Which machine/deep learning algorithm performs the best for detecting zero-day malware?
\item[\bf RQ. 4] Does the batch size, i.e., the size of the next batch of data for training a model, influence the accuracy of predictions?
\item[\bf RQ. 5] Does over-training a machine/deep learning model increase the accuracy of detecting zero-day malware?

\end{description}

\section{State-of-the-Art: Zero-Day Attacks}
\label{sec:background}

Bileg and Dumitras \cite{Bilge2012} conducted a study on zero-day attacks to identify their characteristics before and after disclosure. Based on their study, a zero-day attack lasts approximately for $312$ days and at most 30 months. It was also reported that once these vulnerabilities are revealed, the number of their exploitation increases five times. They also suggested a heuristic approach for identifying zero-day attacks.

Miller \cite{Miller2018} investigated several research questions about efficient machine learning systems for detecting zero-day malware including the application of feature selection techniques for this problem. It was suggested that the best approach for feature selection is using a combination of features but at the same time reducing the number of features based on the classifier type. Miller also suggested that combining classification and clustering techniques and taking advantage of using different classifiers would lead to better results and a safety net against false positives.

Sharma et al.\ \cite{Sharma2018} proposed a prevention strategy, called Distributed Diagnosis System (DDS), based on the context graph strategy. The DDS system consists of three parts:  1) Central Diagnosis System (CDS), 2) Local Diagnosis System (LDS), and 3) Semi Diagnosis System (SDS). The DDS system was developed to guarantee the protection of IoT devices once a possible zero-day vulnerability is identified. Their strategy considers a protocol for sharing sensitive data among IoT devices and other network entities. As a general prevention strategy, if an IoT device is infected, it will be removed from the network before exploiting other entities. 

Alazab et al.\ \cite{Alazab2011} developed a machine learning framework to detect zero-day vulnerability. They trained eight different classifiers which worked based on the API call sequences and frequencies, as the feature set. Their framework first retrieves the API call sequences from the disassembled executable and then updates the signature database based on these API calls and then reports the similarity value. The classification is applied using a supervised learning algorithm with eight classifiers. They achieved an acceptable true positive and false positive rates using Support Vector Machine (SVM) to detect unknown malware samples with obfuscated code.

Comar et al.\ \cite{Comar2013} developed a two-level framework for zero-day malware detection. The framework  consists of six parts: 1) a data capture module, 2) an intrusion detection/prevention system (IDS/IPS), 3) an information storage, 4) the feature extraction and transformation module, 5) a supervised classifier, and 6) an UI portal. Using a class-based profiling technique, their framework was able to separate unknown malware form the known ones and then using network traffic features for their SVM classifier it could detect zero-day malware.

Zho and Pezaros \cite{Zhou2019} evaluated six different machine learning models for zero-day intrusion detection. The train dataset consists of fourteen different intrusions and their test dataset consists of eight different types of zero-day attacks. Using a decision tree classifier, they achieved 96\% accuracy, 5\% false positive rate with acceptable overhead.

Xiao et al.\ \cite{Xiao2018} designed an IoT attack model and discussed different learning techniques for IoT security. The supervised, unsupervised, and reinforcement learning algorithms were employed for different security aspects such as malware detection, authentication, and access control. They provided a few backup security solutions with the machine learning-based security schemes to enable reliable and secure IoT services. They also  recommended some solutions to avoid problems such as oversampling, insufficient training data, and bad feature extraction in the learning methods. 

\section{Experimental Setup}
\label{sec:setup}

\subsection{Dataset and Features}

We obtained the data from the Kaggle Website \cite{ZeroDayDataSet}. According to the hosting Website, the raw dataset had been obtained from the malware security partner of Meraz'18, the annual techno-cultural festival of IIT Bhiali. According to the contributor, the raw dataset included features for both malware and legitimate files. 
The dataset is offered in two pieces: The first piece of data is the training dataset with $138,047$ records of sample data. The sample data itself consists of $96,724$ records whose {\tt legitimate} parameter is labeled as ``0'' and $41,323$ records are labeled as ``1''. The number of features is reported as $55$.  The second part of the data is ``unlabeled'' test data with $88,258$ records of sample data and $55$ features. Since the second part of the data is unlabeled, we only conducted experiments on the first part of the (i.e., the training dataset). 

We ignored three features while building our classifiers namely: {\tt ID}, {\tt Machine}, and {\tt SizeOfOptionalHeader}, since these features appeared to be constant over the data. Table \ref{tab:features} lists the features that were used for building the classifiers (i.e., $52$). As the class label,  the {\tt legitimate} feature was utilized, where  values ``0'' and ``1'' represent the benign and malicious files, respectively.
A cross-validation strategy (i.e., $k$-fold) is utilized to ``validate'' the fitness of the models built. 

\begin{table}[t]
\begin{center}
	\caption{The selected features of the dataset.}
	\rowcolors{2}{gray!30}{white}
	\begin{tabular}{c|c}
		\rowcolor{gray!70}
		\hline 
		\multicolumn{2}{|c}{\bf Features}   \\ 
		\hline 
Characteristics & MajorLinkerVersion \\
MinorLinkerVersion & SizeOfCode \\
SizeOfInitializedData & SizeOfUninitializedData \\
AddressOfEntryPoint & BaseOfCode \\
BaseOfData & ImageBase \\
SectionAlignment & FileAlignment \\
MajorOperatingSystemVersion & MinorOperatingSystemVersion \\
MajorImageVersion & MinorImageVersion \\
MajorSubsystemVersion & MinorSubsystemVersion \\
SizeOfImage & SizeOfHeaders \\
CheckSum & Subsystem \\
DllCharacteristics & SizeOfStackReserve \\
SizeOfStackCommit & SizeOfHeapReserve \\
SizeOfHeapCommit & LoaderFlags \\
NumberOfRvaAndSizes & SectionsNb \\
SectionsMeanEntropy & SectionsMinEntropy \\
SectionsMaxEntropy & SectionsMeanRawsize \\
SectionsMinRawsize & SectionMaxRawsize \\
SectionsMeanVirtualsize & SectionsMinVirtualsize \\
SectionMaxVirtualsize & ImportsNbDLL \\
ImportsNb & ImportsNbOrdinal \\
ExportNb & ResourcesNb \\
ResourcesMeanEntropy & ResourcesMinEntropy \\
ResourcesMaxEntropy & ResourcesMeanSize \\
ResourcesMinSize & ResourcesMaxSize \\
LoadConfigurationSize & VersionInformationSize \\
\hline
\multicolumn{2}{c}{legitimate (class Label)}\\
	\hline 
	\end{tabular} 
	\label{tab:features}
\end{center}
\end{table}

\begin{figure*}
  \includegraphics[width=18cm]{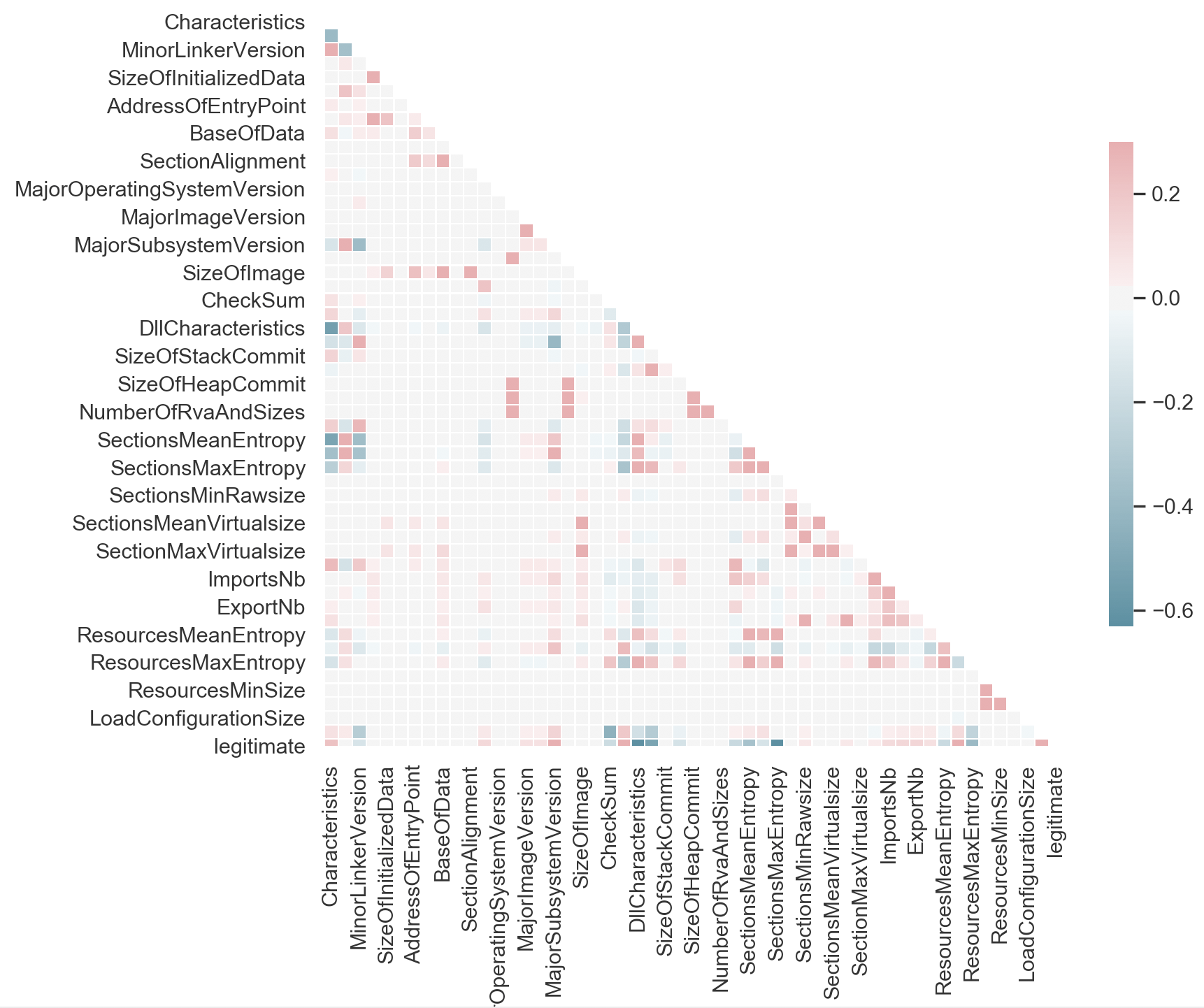}
  \caption{Correlation heat map of the features.}
  \label{fig:heatmapcorr}
\end{figure*}

Figure \ref{fig:heatmapcorr} illustrates the correlation matrix of the features. The correlation values vary in the range of $(-0.6, +0.3)$. Most of the correlation values are close to zero indicating that there is no strong correlation between most of the features.

\subsection{Evaluation Metrics}

The {\tt model\_selection.cross\_val\_score} 
method was used to evaluate the performance of the classifiers studied. The method is offered by the Python {\tt sklearn} package. According to its documentation, the method returns the accuracy of cross validation where accuracy is defined as the fraction of true (i.e., correct) predictions of the model. The accuracy metric can be formally computed as following:

  \vspace*{-0.15in}
\begin{equation}
\begin{split}
Accuracy = & \frac{\#Correct\ Predictions}{\# Total\ Predictions} \\
= & \frac{TP+TN}{TP+ TN + FP + FN}
\end{split}
\end{equation}

Where TP, TN, FP, and FN represent True Positives, True Negatives, False Positive, and False Negatives, respectively.

\subsection{Experimental Setup}
We used the open-source Anaconda with Python $3.7.4$, TensorFlow $1.14.0$, and ran the experiments on a MacBook Pro laptop computer with 2.9 GHz Intel Core i9 processor and 32GB of Ram with 2400 MHz DDR4. 

\section{The Chosen Classifiers}
\label{sec:classifiers}

The research team chose a diverse set of machine and deep learning algorithms grouped into three major classes: 
\begin{itemize}
\item[--] {\bf Conventional Machine Learners} including 1) Gaussian Naive Bayes, 2) Quadratic Discriminant Analysis (QDA), 3) Logistic Regression, 4) AdaBoost, 5) $K$-Nearest Neighbors, 6) Decision Tree, and 6) Random Forest.
\item[--] {\bf Simple Neural Network with a single Layer}, in which the primary algorithm was the Multi-Layer Perceptron (MLP) with a single layer of training. 
\item[--] {\bf Simple Neural Network with Multiple Layers and a Small Value for Epoch}, in which an MLP-based algorithm was utilized with multiple layers and a small number of model fitting.  
\item[--] {\bf Simple Neural Network with Multiple Layers and a Larger Value for Epoch}, in which an MLP was utilized with multiple layers and a large number of training. 
\end{itemize}

The research team developed several Python scripts and utilized relevant packages to carry out the experiments. These machine/deep learning algorithms take in several  parameters that were controlled for the purpose of this study. Table \ref{tab:PythonPKG} lists the Python packages used and the parameters sets.

\begin{table*}[t]
\begin{scriptsize}
	\caption{The Python packages and parameter settings explored in this study.}
	\rowcolors{2}{gray!30}{white}
	\begin{tabular}{p{3.5cm}|l|c|p{4cm}}
		\rowcolor{gray!70}
		\hline 
		\multicolumn{1}{c|}{\bf Classifier} &
		\multicolumn{1}{|c|}{\bf Type} &
		\multicolumn{1}{|c|}{\bf Python Package} &
		\multicolumn{1}{|c}{\bf Parameter Set \& Values} \\
		\hline
		\multicolumn{4}{c}{\bf  Conventional Machine Learning} \\ 
		Adaptive Boosting  & Feature Reduction &
		{\tt AdaBoostClassifier(...)} & 
		{\tt n\_estimators} $n = {50, 100, 200}$\\
		Decision Tree & Tree-based Decision Support & {\tt DecisionTreeClassifier(...)} & Default \\
		Gaussian Naive Bayes & Probabilistic & {\tt GaussianNB(...)} & Default \\
		Linear SVM (LinearSVC) & Supervised Learning & {\tt LinearSVC(...)} & Default \\ 
		Logistic Regression & Logistic function & {\tt LogisticRegression(...)} & Default\\
		$k$ Nearest Neighbors & Unsupervised Learning & {\tt KNeighborsClassifier(...)} & {\tt n\_neighbors} $n=3, 5, 7$\\
		Quadratic Discriminant Analysis & General form of Linear Classifier & {\tt QuadraticDiscriminantAnalysis(...)} & Default \\
		Random Decision Forests & Ensemble Learners & {\tt RandomForestClassifier(...)} & {\tt n\_estimators} $n={10, 50, 100}$\\
		\multicolumn{4}{c}{\bf Neural Networks with Single Layer} \\
		Multi-Layer Perceptron: MLP  & Feed-forward Neural Network & {\tt MLPClassifier(...)} & Number of Neurons \\
		\multicolumn{4}{c}{\bf Neural Networks with Multiple Layers and Small Epoch} \\
		Multi-Layer Perceptron: MLP  & Feed-forward Neural Network & Self-Developed & \\
		\multicolumn{4}{c}{\bf Neural Networks with Multiple Layers and Larger Epoch} \\
		Multi-Layer Perceptron: MLP  & Feed-forward Neural Network & Self-Developed & \\
		\hline
	\end{tabular} 
	\label{tab:PythonPKG}
\end{scriptsize}
\end{table*}


\begin{itemize}
    \item[--] {\tt AdaBoostClassifier()} is 
    a meta-estimator that starts off with fitting a classifier on the original dataset. It then fits additional classifier with different weights on the same dataset to take the classifiers attention towards those instances, which have been classified incorrectly. The 
    {\tt n\_estimators} parameter is the maximum number of estimators to consider when boosting is performed. The remaining parameters were left as default. 
    \item[--] The {\tt DecisionTreeClassifier()} classifier
    is a decision tree classifier whose parameters were set as default. 
    \item[--] {\tt GaussianNB()} performs online updates to model parameters. The default parameter values were used. 
    \item[--] {\tt LinearSVC()} is an efficient implementation of SVC with a linear kernel. The parameters were left as default. The Python {\tt SVC{...}} library, an inefficient implementation of the SVM classifier, did not produce any results even after several hours of executions. 
    \item[--] {\tt LogisticRegression()} An implementation of regular logistic regression. The default values for the parameters were used.
    \item[--] The {\tt KNeighborsClassifier()} classifier is an implementation of the $k$ nearest neighbors clustering. The parameter {\tt n\_neighbors} was used to determine the number of clusters. The remaining parameters set as default. 
    \item[--] {\tt QuadraticDiscriminantAnalysis()} A classifier that employs a quadratic decision boundary to group data. The parameters left as default. 
    \item[--] {\tt RandomForestClassifier()} An ensemble-based classifier that fits a number of decision tree classifiers on different subsets of the  data. The parameter {\tt n\_estimators} is the number of trees in the forest. The remaining parameters left as default. 
    \item[--] {\tt MLPClassifier()} A feed-forwarding neural network. The parameter {\tt hidden\_layer\_sizes} represents the number of neurons in the hidden layer: 
    \begin{itemize}
        \item A call to this library with only one parameter value implies that there is only one layer with the number of neurons specified. {\tt MLPClassifier(100)} implies that there is only one layer with 100 neurons. 
        \item A call to this library with two values implies that there are two hidden layers whose number of neurons is specified. For instance, {\tt MLPClassifier(100, 1)} means that there are two hidden layers each with 100 and 1 neurons, respectively. 
        \item A call to this library with three parameter values implies that there are three hidden layers whose number of neurons is specified as inputs. For instance, {\tt MLPClassifier(100, 50, 1)} means that there are three hidden layers each with 100, 50, and 1 neurons, respectively. 
    \end{itemize}
\end{itemize}

We noticed some differences in the performance when dealing with raw (i.e., non-standardized) versus standardized data. The standardization of data means changing the values to enforce the distribution of standard deviation from the mean to be equal to one. For the comparison purposes and discovering whether standardization affects the performance, we performed the experiments on both non-standardized and standardized data. The Python {\tt StandardScaler(...)} library was used for standardization. 

We performed a stratified $10$-fold cross validation on the dataset. Furthermore, we employed the Python pipeline (i.e., {\tt pipeline(...)}) to sequentially apply a list of transforms and produce a final estimator. The primary purpose of utilizing the pipeline is to add several layers of fitting-assessing procedure. More specifically, the pipeline helped us with stacking two processes: 1) the standardization of the data, and 2) the application of the specified classifier.

\section{Results}
\label{sec:results}

Table \ref{tab:performanceclassifiers} reports the performance of each classifiers The performance is measured in terms of the mean values of accuracy obtained along with their standard deviations. The table also reports the performance achieved by each classifier when their data were standardized or left as raw (non-standardized). The classifiers are ordered with respect to accuracy achieved by each classifier.

\begin{table*}[t]
\begin{center}
	\caption{The performance of classifiers in ascending order of ``Accuracy''  for piped and standardized data ($cv = 10$).}
	\rowcolors{2}{gray!30}{white}
	\begin{tabular}{l|r|r|r|r}
		\rowcolor{gray!70}
		\hline 
		\multicolumn{1}{c|}{\bf } &  \multicolumn{4}{c}{\bf Accuracy \%} \\
		\multicolumn{1}{c|}{\bf } &  \multicolumn{2}{c|}{\bf No Standardized}  & \multicolumn{2}{c}{\bf Piped and Standardized}\\ 
		\multicolumn{1}{c|}{\bf Classifier} &  \multicolumn{1}{c|}{\bf  Mean} &  \multicolumn{1}{c|}{\bf  SD}  & \multicolumn{1}{c|}{\bf  Mean} &  \multicolumn{1}{c}{\bf  SD} \\ 
		& & & \multicolumn{1}{c|}{\bf  (Sorted)} & \\
		\hline 
		\multicolumn{5}{c}{\bf  I) Poorly Performed Learners} \\ 
		Gaussian Naive Bayes & 76.60\% & 39.49\%  & 46.31\% & 0.46\% \\
        Quadratic Discriminant Analysis (QDA) & 46.55\% & 34.29\%  & 49.28\% & 0.45\% \\
        \hline
		\multicolumn{1}{c}{\it  Average} & {\bf \color{blue} 61.57\%} & & {\bf \color{blue} 47.79\%} & \\
		\hline
		\multicolumn{5}{c}{\bf  II) Conventional Machine Learning} \\ 
		Logistic Regression & 30.28\% & 45.20\% & 97.51\% & 0.19\%\\
        Linear SVM (LinearSVC) & 84.51\% & 27.68\% & 97.61\% & 0.16\% \\
        AdaBoost($n = 50$) & 97.57\% & 3.65\%  & 98.78\% & 0.09\%\\
        AdaBoost($n = 100$) & 97.77\% & 3.35\%  & 98.88\% & 0.06\% \\
        AdaBoost($n = 200$) & 97.53\% & 3.87\%  & 98.99\% & 0.08\% \\
        Nearest Neighbors ($k = 7$) & 97.88\% & 2.01\%  & 98.99\% & 0.08\% \\
		Nearest Neighbors ($k = 5$) & {\bf \color{red} 97.89\%} & 2.12\% & 99.07\% & 0.08\% \\
        Nearest Neighbors ($k = 3$) & 97.81\% & 2.45\%  & 99.12\% & 0.06\% \\
	    Decision Tree & 95.20\% & 9.60\%  & 99.24\% & 0.07\%\\
		Random Forest ($n=10$) & 96.85\% & 6.94\%  & {\bf \color{red} 99.45\%} & 0.06\%\\
		Random Forest ($n=100$) & 97.04\% & 6.89\%  & {\bf \color{red} 99.51\%} & 0.08\% \\
		Random Forest ($n=50$) & 97.07\% & 6.66\%  & {\bf \color{red} 99.51\%} & 0.06\% \\
		\hline
		\multicolumn{1}{c}{\it  Average} & {\bf \color{blue} 90.61\%} & & {\bf \color{blue} 98.88\%} & \\
		\hline
		\multicolumn{5}{c}{\bf  III) Simple Neural Network (One Layer)} \\ 
		Multi-Layer Perceptron: MLP ($52$)[$BatchSize=Auto$] & 94.70\% & 4.36\%  &  99.14\% & 0.07\%\\
		Multi-Layer Perceptron: MLP ($100$)[$BatchSize=Auto$] & 95.19\% & 4.55\%  & 99.21\% & 0.11\%\\
		Multi-Layer Perceptron: MLP ($200$)[$BatchSize=Auto$] & 95.52\% & 4.12\% & 99.24\% & 0.09\% \\
		Multi-Layer Perceptron: MLP ($400$)[$BatchSize=Auto$] & 93.72\% & 5.05\% & {\bf \color{red} 99.25\%} & 0.08\% \\
		\hline
		\multicolumn{1}{c}{\it  Average} & {\bf \color{blue} 94.78\%} & & {\bf \color{blue} 99.21\%} & \\
		\hline
		\multicolumn{5}{c}{\bf  IV) Deep Learning (Multiple Layers): $Epoch = 5$} \\
		Multi-Layer Perceptron: MLP (30, 1)[$BatchSize=100$] & 65.01\% & 11.97\% & 98.73\% & 0.09\%\\		
		Multi-Layer Perceptron: MLP (52, 1)[$BatchSize=100$] & 57.13\% & 16.26\% & 98.79\% & 0.07\%\\	
		Multi-Layer Perceptron: MLP (100, 1)[$BatchSize=100$] & 45.04\% & 17.57\% & 98.83\% & 0.07\% \\
		Multi-Layer Perceptron: MLP (30, 1)[$BatchSize=5$] & 65.88\% & 11.99\% & 98.88\% & 0.09\%\\		
		Multi-Layer Perceptron: MLP (52, 1)[$BatchSize=5$]& 58.03\% & 18.26\% & 98.90\% & 0.09\%\\
		Multi-Layer Perceptron: MLP (52, 30, 1)[$BatchSize=100$] & 67.58\% & 15.08\% & 98.90\% & 0.07\% \\
		Multi-Layer Perceptron: MLP (100, 1)[$BatchSize=5$] & 45.04\% & 17.57\% & 98.92\% & 0.10\% \\
		Multi-Layer Perceptron: MLP (100, 50, 1)[$BatchSize=100$] & 50.02\% & 20.05\% & 98.95\% & 0.10\% \\
		Multi-Layer Perceptron: MLP (52, 30, 1)[$BatchSize=5$] & 64.54\% & 18.83\% & 98.95\% & 0.09\%  \\
		Multi-Layer Perceptron: MLP (100, 80, 60, 40, 20, 10, 1)[$BatchSize=5$] & 66.05\% & 12.04\% &  98.97\% & 0.07\% \\
		Multi-Layer Perceptron: MLP (100, 50, 1)[$BatchSize=5$] & 50.02\% & 20.05\% & 98.99\% & 0.10\% \\
		Multi-Layer Perceptron: MLP (100, 80, 60, 40, 20, 10, 1)[$BatchSize=100$] & 70.07\% & 0.00\% & 98.99\% & 0.10\% \\
		\hline
		\multicolumn{1}{c}{\it  Average} & {\bf \color{blue} 58.70\%} & & {\bf \color{blue} 98.90\%}& \\
		\hline
		\multicolumn{5}{c}{\bf  V) Deep Learning (Multiple Layers Different Epochs} \\
		Multi-Layer Perceptron: MLP (100, 50, 1)[$BatchSize=100$],[$Epoch = 50$] & --  & --  & 99.24\% & 0.10\% \\
		Multi-Layer Perceptron: MLP (100, 50, 1)[$BatchSize=100$],[$Epoch = 100$] & --  & --  & 99.28\% & 0.08\% \\
		Multi-Layer Perceptron: MLP (100, 50, 1)[$BatchSize=100$],[$Epoch = 400$] & -- & --  & 99.29\% & 0.06\% \\
		Multi-Layer Perceptron: MLP (100, 50, 1)[$BatchSize=100$],[$Epoch = 200$] & --  & --  & 99.33\% & 0.06\% \\
		\hline
		\multicolumn{1}{c}{\it  Average} & {\bf \color{blue} --} & -- & {\bf \color{blue} 99.28\%}& \\
		\hline
	\end{tabular} 
	\label{tab:performanceclassifiers}
\end{center}
  \vspace*{-0.25in}
\end{table*}


The classifiers are grouped into five classes with respect to their demonstrated performance: I) Poorly performed classifiers, II) Conventional Machine Learning classifiers with good performance, III) Simple Neural Network with Single Hidden Layer, IV) Deep Learners with small Epochs, and V) Deep Learners with larger Epochs.

\subsection{RQ. 1: The Influence of Standardization?} 

A quick glance at the accuracy achieved by applying each classifier on non-standardized and standardized forms of data shows that (except a couple of cases (Gaussian, Naive Bayes and Quadratic Discriminant Analysis (QDA))), the classifiers demonstrated an improved accuracy when the given data are standardized. It was also observed that it takes much smaller amount of time and effort to train models with standardized data; whereas, when non-standardized data are fed into the classifiers, it takes considerable amount of time for training and fitting the models. 

The average mean values for accuracy obtained for different classes of classifiers when non-standardized data are modeled are $61.57\%$, $90.61\%$, $94.78\%$, and $58.70\%$ for Poorly performed classifiers, conventional machine learners, Simple Neural Networks with only one hidden layer, and Deep learning models, respectively. Whereas, the calculated mean values for accuracy when standardized data are used are $47.79\%$, $98.88\%$, $99.21\%$, and $98.90\%$ for Poorly performed classifiers, conventional machine learners, Simple Neural Networks with only one hidden layer, and Deep learning models, respectively. Excluding the poorly performed classes of classifiers, we observe that standardization help in improving accuracy by $98.88 - 90.61 = 8.27\%$, $99.21 - 94.78 = 4.43\%$, and $98.90 - 58.70 = 40.2\%$, for conventional machine learners, Simple Neural Networks with only one hidden layer and Deep learning models, respectively. It was also observed that the standard deviations calculated for accuracy were much higher for non-standardized data compared to standardized data. The primary reason might be due to the fact that larger and wider scale of numerical raw values were used in model fitting. 

\subsection{RQ. 2: Achieving 100\% Accuracy?} 

The research team tested different classifiers with different tuning parameters with the goal of achieving the mean value of $100\%$ for accuracy. It turned out that achieving such a high accuracy on model fitting and prediction was infeasible. While some of the classifiers demonstrated very high accuracy and thus promising results, it seems that building a model that reduces the false positive and false negative ratios to zero is very difficult and hence there are some penalties with missing such cases for detecting zero-day malware. However, it is also possible that building a perfect model with $100\%$ accuracy may imply that the model is overfitted and may perform poorly for classifying unseen data. To avoid such overfitting problem, we employed a $10$-fold cross-validation to optimize the models and that might explain why it was infeasible to build a perfect model with zero false positive and false negative values. 

\subsection{RQ. 3: The Best Classifier?} 

As Table \ref{tab:performanceclassifiers} shows, the random forest was the best classifier with outstanding performance of achieving $99.51\%$ on average for accuracy, followed by decision tree achieving $99.24\%$. The simple neural networks with a single hidden layer also performed very well. On average, they achieved $99.21\%$ on accuracy. The number of neurons on the singleton hidden layer seems to have some light impacts on accuracy. For instance, an MLP model with only one hidden layer and 50 neurons achieved $99.14\%$ accuracy; whereas, increasing the number of neurons to 100, 200, and 400 increased the accuracy to $99.21\%$, $99.24\%$, and $99.25\%$, respectively. Since there are some computational costs associated with the number of neurons on the layer, and given the slights improvement on the observed accuracy, the question is whether a more complex model is worthy to be built or a simpler model with slightly lower accuracy would be sufficient for the prediction. The choice of this trade-off totally depends on the application domain. As a special case, detecting zero-day malware is an important and critical task and thus increasing the accuracy as much as possible is indeed needed regardless of the cost. 

\subsection{RQ. 4: The Influence of Batch Size in Deep Learning?} 

The authors controlled the batch size parameters for deep learning classifiers with multiple layers. According to our observations: the smaller batch size is, the better/fitter the model will be. For instance, a neural network with two hidden layers each with 10 and 1 neurons but with batch size of 100 and 5 achieved the accuracy of $98.73\%$ and $98.88\%$, respectively. The improvement seems to be very small. On the other hand, training with smaller batch size appeared to take more computation time and thus more expensive than training a model with a larger batch size.

It was also observed that for deep learning-based approaches with multiple layers: the larger and deeper the model is, the better and fitter the classification will be. For instance, a model with two layers each having 30 and 1 neurons and batch size of 100 achieved the accuracy of $98.73\%$. Whereas, a deeper model with seven layers each having 100, 80, 60, 40, 20, and 1 neurons and with the batch size equal to 100 achieved the accuracy of $98.99\%$. However, the improvement is very small. 

\subsection{RQ. 5: The Influence of Epochs in Deep Learning?} 

\begin{figure}
  \includegraphics[width=8.5cm]{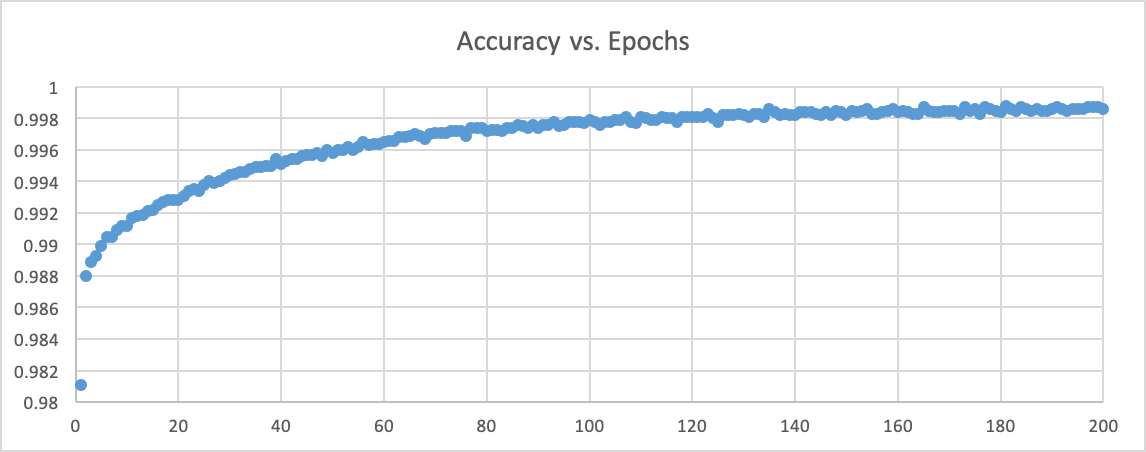}
  \caption{Accuracy vs. Epochs.}
  \label{fig:auuvsepoch}
    \vspace*{-0.25in}
\end{figure}

The authors performed a systematic analysis on the influence of the number of iterations needed to train the classifiers. It was observed that: the greater the number of epochs is, the more accurate the model will be. For instance, an MLP model $(100, 50, 1)$ with $Epochs = 50$ achieved $99.24\%$ accuracy; whereas, an increase of Epochs to 200 enhanced the accuracy to $99.33\%$. This observation might indicate that a smaller number of epochs might be sufficient to learn the key and significant features of the data and thus by adding more rounds of training stages, the model will not learn anything further (i.e., all features are already learned). As an example, Figure \ref{fig:auuvsepoch} illustrates the improvement of accuracy over epochs for the model $MLP(100, 50, 1)[BatchSize=100][Epoch=200]$.

\section{Discussion}
\label{sec:discussion}

\subsection{Standardization Is Important for Classification}

According to our results, standardization is critical for classification. The primary reason might be because of computational expenses involved in dealing with large numbers and thus with higher standard variations. Some of these classifiers utilize a distance metric (e.g., Euclidean distance) where the square roots of the sum of the squared differences between the observations are calculated for clustering the data items. As a result, to accommodate such expensive computation when larger values are provided as data, the demands for computational needs will be increased. Hence, since a larger standard deviation will affect the accuracy of the prediction.

\subsection{Feature Reductions and Parameters Tuning}
The authors tuned several parameters of the classifiers. Furthermore, they studied several machine/deep learning algorithms. These learners perform the task of classifications differently. For instance, one utilizes hyperplanes to create clusters (e.g., SVM); whereas, some other uses ensemble learning and take the majority votes to decide (e.g., random forest). Moreover, some of these techniques apply feature reductions and thus tune the parameters and build a model; whereas, some other more advanced algorithms try to take into account all features and then through deeper analysis adjust their contributions and weights to the final model. A potential drawback of utilizing all features in the computation is the overfitting problem and thus the model may suffer from being tightly coupled to the seen data and thus unable to perform well for unseen data. It also may cause adding noisy features into the computations. On the other hand, reducing features may cause the problem of missing important relationships between parameters. The choice of feature reduction depends on the type of dataset and is an important decision and should be handled with additional care. 

\subsection{Conventional vs. Deep Learning Classifiers}
The authors expected to observe much better performance from deep learning-based algorithms. While these deep learners performed very well, surprisingly, some of the conventional machine learning classifiers performed comparatively similar or even better. Given the lower cost of training associated with the conventional machine learning algorithms and at the same time a considerably greater cost for training deep classifiers, the conventional machine learning algorithms might be even a better choice compared to the deep learning-based algorithms. 

The deep learning-based classifiers demonstrate a consistent improvement achieved by building larger models and additional training. However, a simple random forest algorithm still outperforms even larger deep learning-based classifiers with additional training. For instance, the performance demonstrated by Random Forest (i.e., $99.51\%$) and the best performance achieved by deep learning (i.e., $99.33\%$) after 200 epochs with three hidden layers is remarkable.

\subsection{Deep or Deeper Classifiers?}

According to our results, larger and deeper classifiers tend to perform better and they build a more accurate model. However, the improvement does not seem to be significant. A simpler deep learning classifier (e.g., $MLP(30, 1)$ with batch $size = 100$ and accuracy of $98.73\%$) might perform comparatively very similar to a deeper and larger classifier (e.g., $MLP(100, 80, 60, 40, 20,. 10, 1)$ with batch $size = 100$ and accuracy of $98.99\%$. Hence, the choice of the depth of deep classifiers depends on the desired level of accuracy. 

\section{Conclusions and Future Work}
\label{sec:conclusion}

This paper empirically explored whether machine and deep learning classifiers are effective in detecting zero-day malware. Addressing such a question is important from security perspective because zero-day malware are unknown applications and thus there might not be any malicious signature similar to their patterns. We empirically compared a good number of well-known conventional machine learning and deep learning classifiers and observed that some of the conventional machine learning algorithms (e.g., random forests) perform very well in comparison with their deep learning-based counterparts. This result implies that some of the conventional and deep learning-based approaches are good classifiers for detecting zero-day malware. However, even though they achieve very high accuracy (e.g., $99.51\%$), these algorithms never achieve a $100\%$ accuracy and thus these classifiers might slightly misclassify some of the  zero-day malware. 

This paper focused on measuring the accuracy of classifiers using a $10$-fold cross validation. It is important to carry out additional experiments and measure precision, recall, accuracy, and F1 measures all together along with the ROC measure for these classifiers and capture the exact values for false positive and false negative. It is also important to replicate the experiments reported in this paper with some other datasets and perform a meta analysis \cite{KakarlaMN11} to have a better insights about the machine leaning algorithms and their classification performances. Furthermore, given the outstanding performance of random forest, it would be interesting to observe whether ensemble-based deep learning classifiers perform better than other classifiers. It is also an interesting question to investigate whether evidence theory \cite{ChatterjeeND18}, uncertainty reasoning \cite{SartoliN16a}, or control-theoretical approaches and decision-based processes \cite{ZhengN18} can be utilized in accordance with learning algorithms to detect zero-day vulnerabilities.

\section*{Acknowledgment}
This work is supported in part by National Science Foundation (NSF) under the grants 1821560 and 1723765.
\bibliographystyle{IEEEtran}
\bibliography{References}

\end{document}